\documentclass[final]{vgtc}                          

\graphicspath{{figures/}{pictures/}{images/}{./}} 

\usepackage{times}                     

\usepackage{tabu}                      
\usepackage{booktabs}                  
\usepackage{lipsum}                    
\usepackage{mwe}                       

\usepackage{mathptmx}                  

\onlineid{0}

\vgtccategory{Research}

\vgtcinsertpkg

\title{Evaluating Transferable Emotion Expressions for \\ Zoomorphic Social Robots using VR Prototyping}

\author{Shaun Macdonald
\thanks{e-mail: shaun.macdonald.research@gmail.com}\\ %
         \scriptsize University of Glasgow %
\and Robin Bretin
\thanks{e-mail: robin.bretin@gmail.com}\\ %
         \scriptsize University of Glasgow %
\and Salma ElSayed
\thanks{e-mail: s.elsayed@abertay.ac.uk }\\ %
         \scriptsize Abertay University}

\abstract{%
Zoomorphic robots have the potential to offer companionship and well-being as accessible, low-maintenance alternatives to pet ownership.
Many such robots, however, feature limited emotional expression, restricting their potential for rich affective relationships with everyday domestic users. 
Additionally, exploring this design space using hardware prototyping is obstructed by physical and logistical constraints.
We leveraged virtual reality rapid prototyping with passive haptic interaction to conduct a broad mixed-methods evaluation of emotion expression modalities and participatory prototyping of multimodal expressions.
We found differences in recognisability, effectiveness and user empathy between modalities while highlighting the importance of facial expressions and the benefits of combining animal-like and unambiguous modalities. We use our findings to inform promising directions for the affective zoomorphic robot design and potential implementations via hardware modification or augmented reality, then discuss how VR prototyping makes this field more accessible to designers and researchers.
}


\begin{document}


\firstsection{Introduction}

\maketitle

Zoomorphic robots are an animal-like subset of social robots which have the potential to simulate positive human-animal interactions.
This has been utilised in settings such as hospitals and care homes~\cite{Shibata2009, Jeong2018, Wada2007, Hudson2020} with receptive elderly or young users and where animals may be unavailable.
In many cases, however, the affective expressiveness and interactions of such robots can be limited, restricting their ability to foster engaging pet-like emotional interactions outside of these settings with everyday domestic users ~\cite{Katsuno2022, Sefidgar2016}; rather they rely on personification~\cite{Young2011a, Katsuno2022} and there have been calls to explore richer evolving relationships in Human-Robot Interaction (HRI)~\cite{Tanevska2020}.
It is not yet clear how zoomorphic robots should best express internal emotional states clearly while fostering empathetic relationships with users.
\par
Commercial zoomorphic robots, such as Paro~\cite{Paro}  or Joy for All Cats~\cite{AgelessInnovationLLC}, have been used to promote emotional well-being in care settings~\cite{Shibata2009, Chang2013} and facilitate pet-like interactions~\cite{Hudson2020}, with emotive movement and sound.
When compared to social robots with built-in screens, however, investigating affective expression for zoomorphic robots can be challenging, as adding display modalities requires modifying their non-modular hardware without disrupting their zoomorphic features.
For example, robots with faces or tails which are static or cannot be directly controlled would require obstructive modification to explore these modalities.
The physical constraints of augmenting robots with multiple expression modalities, alongside the required technical expertise, cost, time-consuming design and implementation, also restricts research using custom-made robots ~\cite{Loffler2018, Song2017}.
This has resulted in evaluations with simple, abstract robots displaying a small selection of modalities, such as light, sound and motion~\cite{Loffler2018, Song2017, Singh2013}.
We address these limitations using Virtual Reality (VR) rapid prototyping, enabling exploration of this design space beyond the constraints of physical prototyping~\cite{Katsuno2022}, while retaining physical interaction with the robot.
This approach has been advocated for~\cite{Suzuki2022, Meyer2021} and used in other fields, such as prototyping smart objects or authentication~\cite{Mathis2021, Voit2019}, but we are the first to apply it to zoomorphic robots.
\par
\begin{figure}[h!]
    \centering
  \includegraphics[width=.99\linewidth]{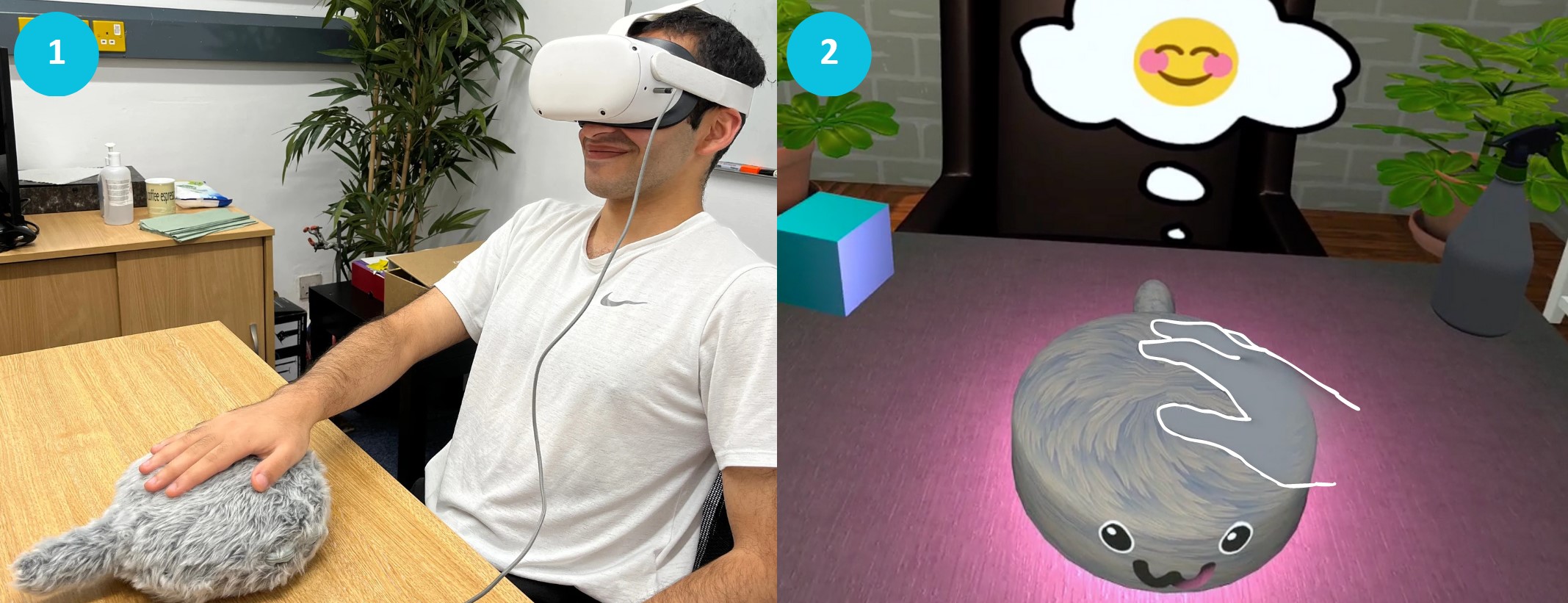}
  \caption{(1) A user touching a zoomorphic robot, (2) experiencing a multimodal emotional expression prototyped in virtual reality.}
  \label{fig:teaser}
\end{figure}
We present the largest to-date comparative evaluation of robot emotion expressions, using interactive VR prototyping and mixed methods analysis.
The robot used, Petit Qoobo~\cite{petitQoobo}, expresses no distinct emotions and has an abstract form factor, making it an ideal testbed for transferable emotion expression modalities.
A VR twin was overlaid onto the physical robot, to retain passive haptic feedback  (Fig. \ref{fig:teaser}), and was augmented with seven modalities: \textit{light}, \textit{sound}, \textit{text}, \textit{text-to-speech (TTS)}, \textit{emoji}, \textit{facial expressions}, and the robot's existing modality: \textit{tail movements}.
For each expression, participants attempted to identify the emotion, then rated how effective it was and how empathetic they felt. 
Participants were then able to prototype their own multimodal expressions in the virtual environment before they gave qualitative feedback regarding their modality preferences, animal experience and prototyping rationale.
\par

Overall, expressions were recognised 71\% of the time, with three modalities, \textit{text}, \textit{emoji} and \textit{speech}, recognised over 90\% of the time. \textit{Facial} expressions and \textit{emoji} were most effective at driving emotional connection. 
When participants crafted their own multimodal expressions, \textit{face}, \textit{tail}, and \textit{coloured light} were most used, with modalities that evoked animal interaction and facilitated emotional connection (e.g., \textit{face} or \textit{tail}) being prioritised, presented with less-lifelike modes (e.g., \textit{light} or \textit{text}) to lend clarity or emphasis.
Our findings reveal user priorities for affective expression by zoomorphic robots, informing designers when taking the next step to prototype real robots.
We provide modality recommendations and possible implementations via either hardware modification or augmented reality (AR).
Finally, we discuss how VR prototyping could enable fast and accessible exploration of affective robot design.
\par
\textbf{Contribution Statement: }The primary contribution of this work is a broad user evaluation of seven transferable emotion expression modalities for zoomorphic robots, facilitated by a novel use case of VR prototyping to enable rapid exploration of affective robot design. 
We leverage our findings to offer guidance for zoomorphic robot emotion expression, future work directions and possible implementations using physical hardware or AR.

\section{Related Work}

\subsection{Animal Companionship and Zoomorphic Robots}

\begin{figure*} 
    \centering
        \includegraphics[width=0.19\linewidth]{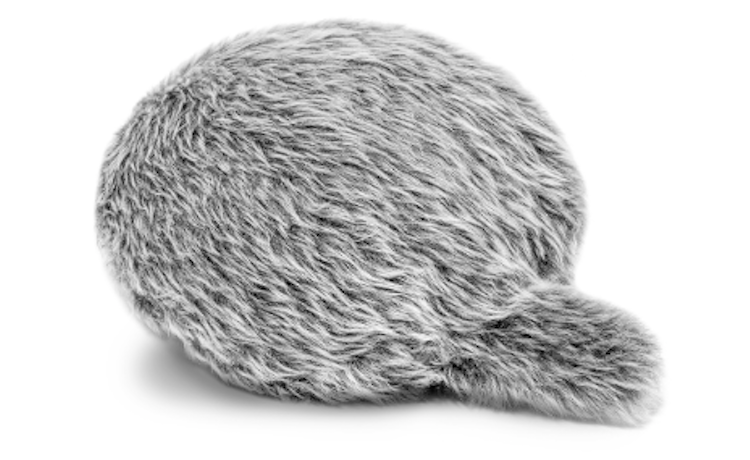}
        \hfill
        \includegraphics[width=0.19\linewidth]{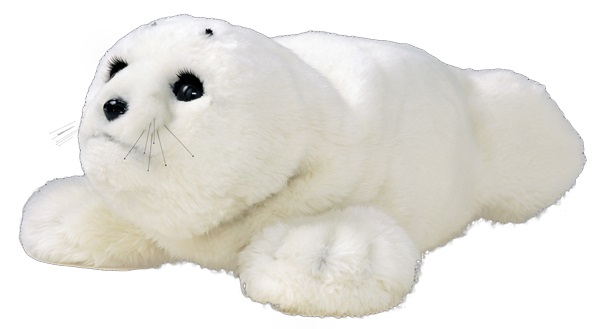}
    \hfill
        \includegraphics[width=0.19\linewidth]{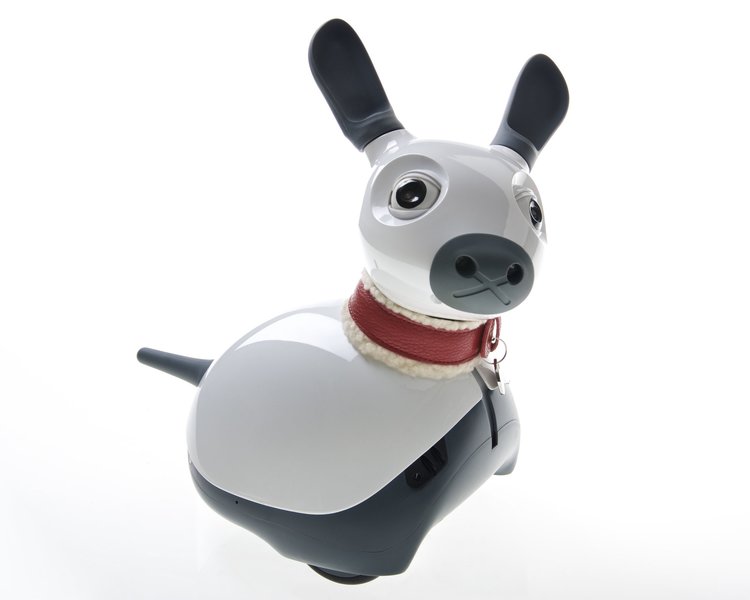}
          \hfill
        \includegraphics[width=0.19\linewidth]{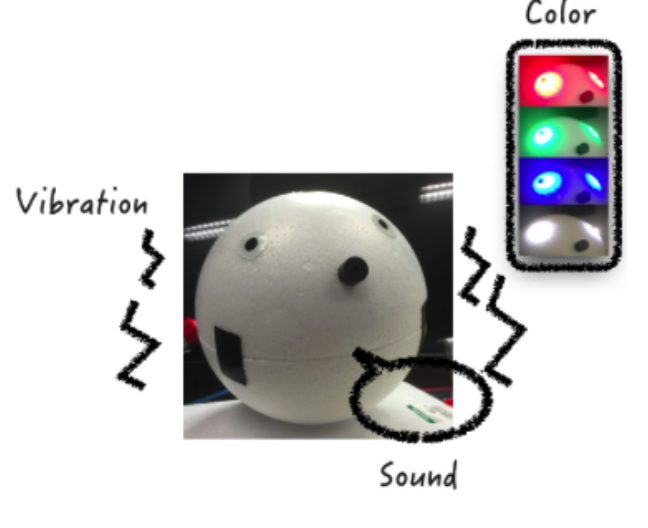}
          \hfill
        \includegraphics[width=0.19\linewidth]{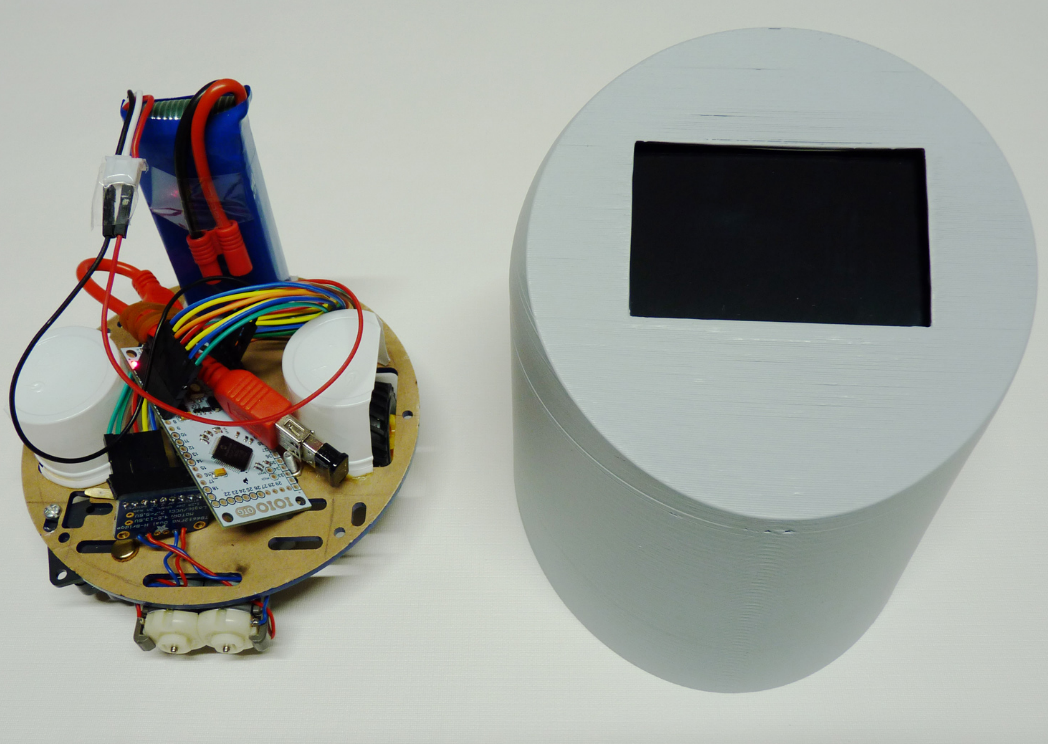} 
  \caption{Zoomorphic and multimodal expressive robots. Left to right: Petit Qoobo, Paro, Miro~\cite{Ghafurian2022}, Maru~\cite{Song2017}, robot built by Löffler et al.~\cite{Loffler2018}.}
  \label{ZoomorphicRobots} 
\end{figure*}

The development of zoomorphic robots is motivated by human desire for animal companionship~\cite{petsUK} and its effects.
Animal-Assisted Therapy (AAT) has shown significant benefits to anxiety, stress and well-being in clinical and non-clinical settings~\cite{Barker1998, Odendaal2000, Nimer2007, Polheber2014, Young2020}.
Animal companionship is, however, not always available or advisable due to hygiene, logistical or financial restrictions \cite{pomba2017public, schwarz2007animal}, motivating exploration of zoomorphic robots as substitutes.
Interactions with the robot dog AIBO have been found to partially emulate the social stimulation a real dog provides for dementia patients~\cite{Kramer2009}, while robots like \textit{Paro} (Fig. \ref{ZoomorphicRobots}) and Joy For All's companion cat have facilitated interpersonal social interaction, companionship, engagement and pet-like interactions with children, elderly users, and in care settings~\cite{Wada2007, Chang2013, Hudson2020, Ihamaki2021, Jeong2018}.
Finally, animal-like interactions have been shown to make participants feel calmer~\cite{Sefidgar2016, borgstedt_soothing_2024} and Bradwell et al.~\cite{Bradwell2021} found familiar zoomorphic designs were most acceptable and avoided fear responses in social care contexts.
\par
There are concerns when moving zoomorphic robots from care settings - with receptive users and a lack of animal interaction - to the home.
They struggle to overcome the “3-month barrier” of engagement, as shallow, repetitive interactions require users to use their imagination to imbue these robots with intimacy~\cite{Berkel2022, Katsuno2022}.
Others have expressed concerns about the impact of companion robots on elderly users~\cite{Sharkey2014}, while users worry about the cost of adoption~\cite{Bradwell2020}.
Shibata et al.~\cite{Shibata2009} found most \textit{Paro} owners used the robot as a substitute as they could no longer care for pets.
This highlights the benefit of zoomorphic robots for users who are unable or unwilling to own pets but would benefit from animal-like interaction, should the gap between robot and pet shrink.
We work to bridge this gap via a broad comparative evaluation of zoomorphic robot emotion expressions, informing the design of expressive interactions.

\subsection{Exploration of Emotion Conveyance in HRI}
\label{bg:emotion conveynce}

This section reviews zoomorphic robot emotion expression to illustrate how we further the field.
First, a model is required to delineate the emotions the robot expresses.
Most affective work on social and zoomorphic robots uses Ekman's model of six basic emotions~\cite{Ekman1992}: happiness, anger, disgust, fear, sadness and surprise~\cite{Cavallo2018}, which we adopted to allow maximum applicability to the field.
Another common method maps emotions to two axes: valence (pleasant to unpleasant) and physical arousal, intersecting at a central origin~\cite{Russell1980}. 
These models are compatible, as basic emotions can be mapped to this model, such as happiness (high valence/arousal), sadness (low valence/arousal) and anger (low valence/high arousal).
\par
Prior work has explored various approaches to zoomorphic emotion expression.
Yohanan et al.'s \textit{Haptic Creature}~\cite{Yohanan2011} used naturalistic modes, including movable ears, vibrotactile purring and expanding lungs, to communicate arousal and valence.
Singh et al.~\cite{Singh2013} investigated a dog-like tail to augment a vacuum robot, finding that varying tail height, direction, speed and wag distance impacted perceived valence and arousal.
Animal interaction has also influenced other robots, as Gacsi et al.~\cite{Gacsi2016} created emotion expressions for an anthropomorphic robot, derived from dog behaviours.

A survey of robot social cues~\cite{Nocentini2019} found most employed one to three modalities, including movement, faces, speech, light and tactile cues. Research has evaluated these modalities in isolation and combination. 
Song et al.~\cite{Song2017} addressed the practical difficulties of comparing multiple modalities by building a simple robot with three modalities (light, sounds, vibration) to express relaxation, happiness, sadness and anger. Happiness was difficult to communicate, but effective expressions were found for sadness (blue light, gentle vibration, falling beeps) and anger (red light, intense vibration, rising beeps). 
Löffler et al.~\cite{Loffler2018} also built a simple robot to display three modalities (light, motion, beeping) to express joy, sadness, scared and anger.
\textit{Emotibots} built by Thompson et al.~\cite{Thompson2024} explored three modes (coloured light, vibration, sound) but without evaluation.  
Finally, Ghafurian et al.~\cite{Ghafurian2022} designed multimodal emotion expressions for the zoomorphic robot \textit{Miro}~\cite{miro}, although most expressions were specific to \textit{Miro}'s articulation capabilities and not transferable, nor were they compared.
We extend prior work by using VR to (1) prototype a broader set of transferable expression modalities onto an existing zoomorphic robot and (2) allow participants to experience, evaluate, and prototype their own multimodal expressions without obstructive modification of the robot.

\subsection{Extended Reality for Prototyping and HRI}
\label{prioruse}

There is precedent for using extended reality (XR) technologies, encompassing VR, AR and mixed reality, to evaluate affective displays.
Social psychology research has leveraged VR to investigate interpersonal emotional expression~\cite{rapuano2020effect}, while others explored prototyping social expressions for drones using VR~\cite{10.1145/3125739.3125774, 9514453}.
While VR has not yet been used in zoomorphic robot expression design, prior work has investigated digital prototyping of robotic affective design.
This includes screen-based desktop approaches, such as using a 3D model to prototype robot facial expressions before physical implementation~\cite{Saldien2010}, or augmenting a vacuum robot with cartoon faces and flourishes via a webcam feed~\cite{Young2007}.
Prior work has also utilised AR to augment robots with new affective capabilities~\cite{Bassyouni2021}, although not with zoomorphic robots.
Groechel et al.~\cite{Groechel2019} used AR to project expressive arms onto a simple social robot, finding user attitudes improved but not exploring emotional expression. 
We build upon these approaches with a novel methodology and focus: applying immersive VR to prototype and evaluate transferable expression modalities for zoomorphic robots, addressing calls from the field to ``re-imagine robot design without physical constraints'' using ``naturalistic'' and ``hand-tracked interactions''~\cite{Suzuki2022}.
\par
XR has also facilitated the prototyping and evaluation of complex systems to inform real-world development in other fields.
VR prototyping has been used to study interactions with public displays~\cite{Makela2020} or explore novel authentication schemes~\cite{Mathis2021}, with evaluation results found applicable to real-world use.
Similarly, VR has facilitated spatial knowledge acquisition by simulating pedestrian navigation~\cite{Savino2019} and, in a comparative evaluation of smart objects using \textit{in-situ}, VR, AR, lab and online approaches,  Voit et al.~\cite{Voit2019} found VR and \textit{in-situ} methods yielded similar results.
Our approach aligns with Meyer et al.~\cite{Meyer2021}, who used mixed reality to explore design space for products that are otherwise complex to develop.

\section{Methodology and Study Design}


\subsection{Study Design}
\label{studydesign}

We conducted a comparative evaluation of emotional expression modalities for zoomorphic robots via a within-subjects design with two independent variables: (1) expression modality and (2) emotion expressed. 
We chose the five emotions from Ekman's model that most commonly appeared in prior work~\cite{Ekman1992}: happy, sad, angry, scared and surprised.
The choice and design of modalities are discussed in Sec. \ref{sec:modality_design}.
Three dependent variables were measured: emotion recognition accuracy,  perceived expression effectiveness and self-reported empathy toward the robot. 
Preconceptions were measured using the Negative Attitude Towards Robots Scale (NARS)~\cite{Nomura2006}.
Qualitative survey and interview data were collected and analysed using thematic analysis to understand participant rationale (Sec. \ref{sec:QualitativeAnalysis}).
The study received IRB approval.
\par
We leveraged a VR prototyping approach to allow new modalities and the existing tail to be rapidly prototyped, controlled and compared without obstructive modification of the robot.
There is necessarily, however, a gap in user experience between a VR twin and a real robot.
We took steps to assess and mitigate this gap: participants were familiarised with the robot, employing passive haptic feedback and immersive features in the virtual environment (VE) (Sec. \ref{VE}).
We administered the Igroup Presence Questionnaire (IPQ)~\cite{schubert2001igroup} and System Usability Scale (SUS)~\cite{SUS} to assess if a lack of presence or usability could negatively bias participants.


\subsection{Participants }We recruited 28 participants (18 female, 10 male) over 17 years old, (mean age = 26.6, $\sigma$ = 4.95, range: 21-40) with unimpeded vision, hearing and use of their hands via institution email channels. 
Participants indicated prior experience with VR and social robots. 
8 had extensive VR experience, 9 had used it `a few times', 6 used it `once or twice' and 5 had no experience.
20 participants had no experience with social robots, 7 did, and 1 was unsure.

\subsection{Apparatus}

The study was conducted in a small room with the robot, \textit{Petit Qoobo}, on a table in front of the seated participant (Fig. \ref{fig:teaser}). 
The robot has animal-like fur, a soft static heartbeat and a moving tail. It does not, however, express distinct internal emotional states like other designs \cite{Song2017, Loffler2018, Ghafurian2022}. 
Instead, it displays different tail motions that users may interpret with affective meaning.
Using a commercial zoomorphic robot with limited emotional expression, allowed scope for affective augmentation, and its abstract form factor and single expression modality (tail movement) made it simple to create a digital twin and augment it with many new modalities without clashing with existing physical features.
For VR portions of the study, a Meta Quest 2 with a glasses-spacer was used.
The VE was run in real-time with Unity Editor V2021.3.13f1 on a Windows 10 Pro Alienware i7 2021 laptop. 
A MacBook was used to administer the information sheet, consent form and surveys via Qualtrics~\cite{qualtrics}.
Interviews were voice-recorded with consent using an iPhone 13.

\begin{figure}[h!]
     \centering
     \includegraphics[width=.47\textwidth]{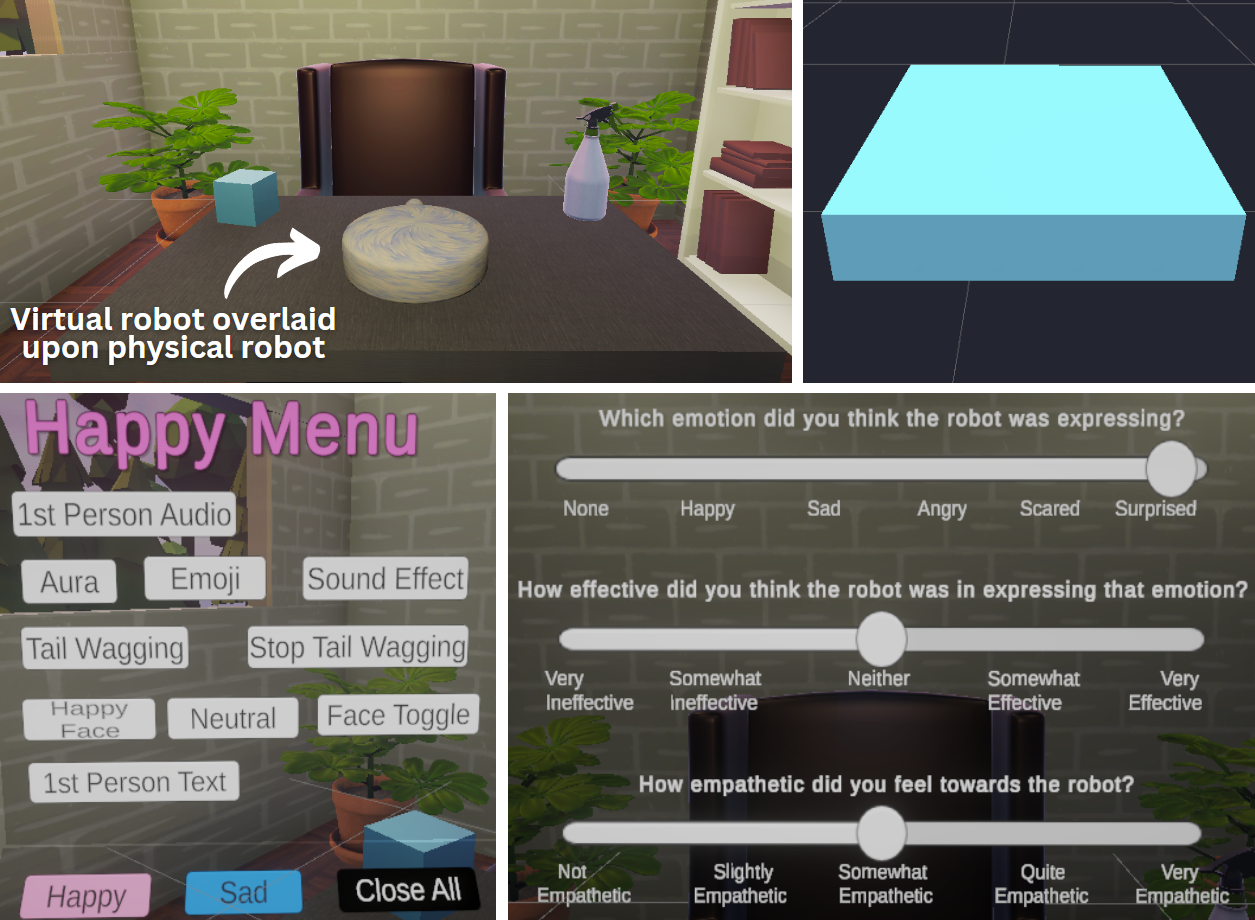}
     \caption{The VE (top left), calibration cuboid (top right), participatory prototyping interface (bottom left) and rating scales (bottom right).}
     \label{fig:VE}   
\end{figure}

\subsection{Virtual Environment Design and Interaction}
\label{VE}

The VE was designed to resemble the study room with a table, two chairs, and similar items placed on the table and around the room (Fig. \ref{fig:VE}).
Environmental details were added to heighten comfort and immersion~\cite{WonImmersive2021}, including a window view of trees and ambient sounds of traffic and birdsong. 
The virtual robot was placed in the same location as its physical counterpart, facilitating passive haptic feedback.
To achieve this, the researcher guided the participant through a calibration scene placing the participant's hand on each side of the physical robot to align it with a virtual cuboid (Fig. \ref{fig:VE}). 
In the VE participants used their finger to interact with UIs to navigate between conditions, rate expressions, or prototype multimodal expressions, facilitated by Quest 2 hand-tracking.
Emotional expressions were triggered by having the participants lay their hand on the robot in an undirected manner, to more closely mirror typical interaction with zoomorphic robots and increase immersion.

\subsection{Study Procedure}

\begin{figure}[h!]
    \centering
        \includegraphics[width=.48\textwidth]{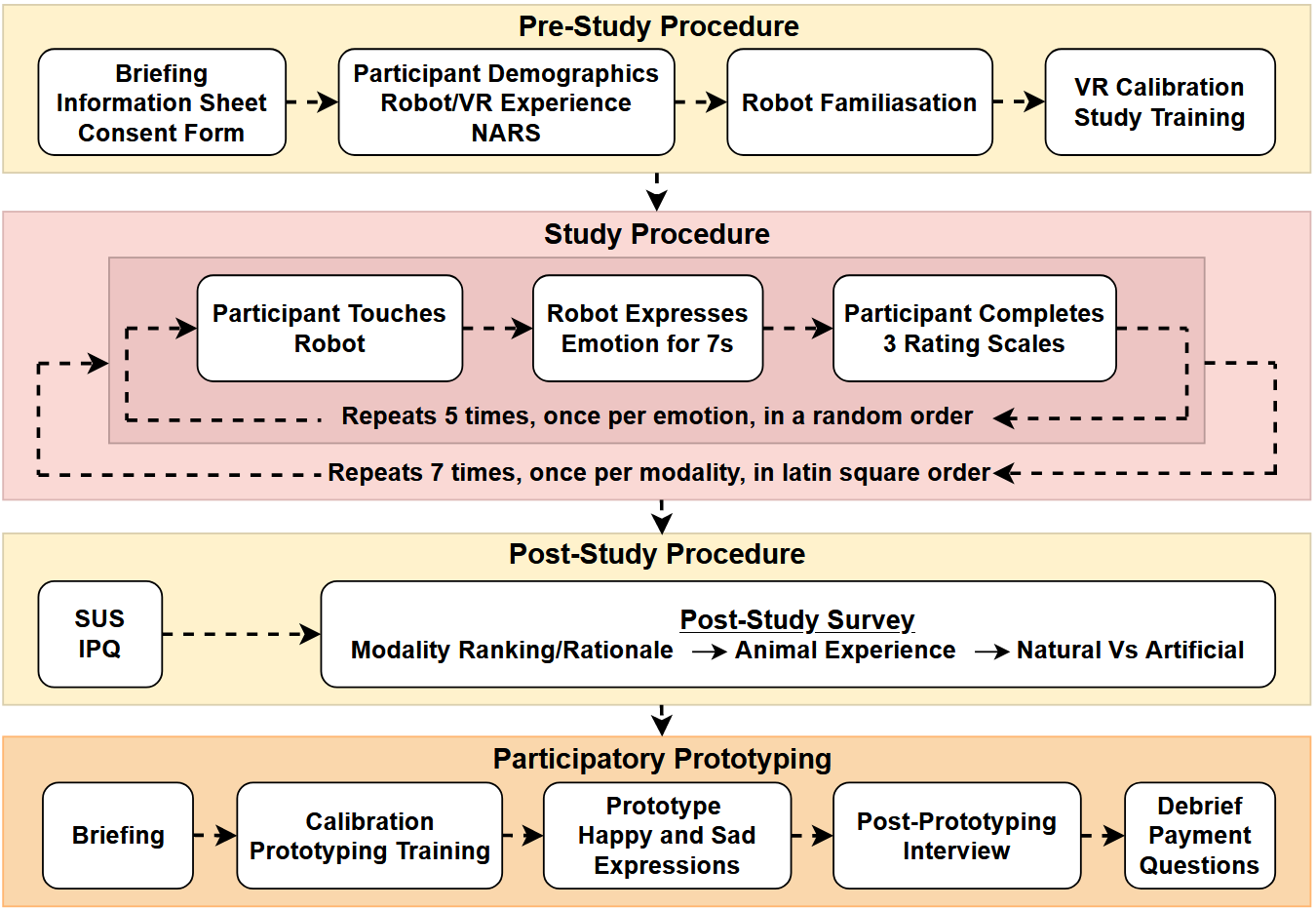}
        \caption{Diagram showing study procedure, divided into four stages: pre-study, study, post-study and participatory prototyping.}
     \label{fig:Procedure}   
\end{figure}

The four-stage study procedure (Fig. \ref{fig:Procedure}) took between 50-60 minutes. Participants were compensated with a £10 Amazon voucher.

\begin{table*}[h!]
\centering
\caption{Details for how each emotion was expressed per each modality. Design rationale is discussed in Sec. \ref{sec:modality_design}.}
\resizebox{.8\textwidth}{!}{%
\begin{tabular}{@{}|c|c|c|c|c|c|c|@{}}
\toprule
\textbf{Mode} & \textbf{Happy} & \textbf{Sad} & \textbf{Angry} & \textbf{Scared} & \textbf{Surprised} \\ \midrule
\hline
\textbf{Face} & \includegraphics[width = 3 cm, height = 1.2 cm]{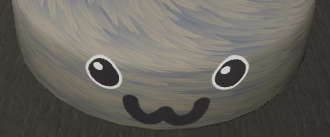} & \includegraphics[width = 3 cm, height = 1.2 cm]{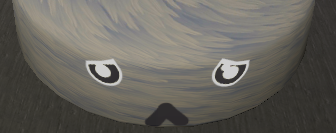} & \includegraphics[width = 3 cm, height = 1.2 cm]{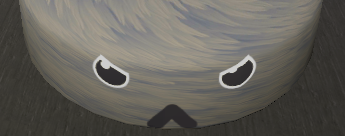} & \includegraphics[width = 3cm, height = 1.2 cm]{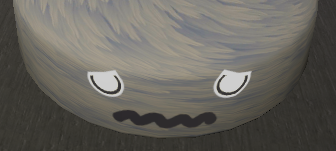} & \includegraphics[width = 3 cm, height = 1.2 cm]{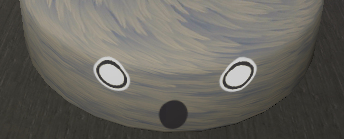} \\ \midrule
\textbf{Tail} & \begin{tabular}[c]{@{}c@{}}`Wide Upward Wags'\\ Width(70$^{\circ}$), Raised (15$^{\circ}$) \\ Speed:20$^{\circ}$/s, Horizontal Wag \end{tabular} & 
\begin{tabular}[c]{@{}c@{}}
'Low Slow Bobs'\\ Width(28$^{\circ}$), Lowered(10$^{\circ}$)\\ Speed:6$^{\circ}$/s, Vertical Wag\end{tabular} & 
\begin{tabular}[c]{@{}c@{}}`Fast Upward Bobs'\\ Width(10$^{\circ}$), Raised(25$^{\circ}$) \\ Speed:100$^{\circ}$/s, Vertical\end{tabular} & 
\begin{tabular}[c]{@{}c@{}}`Low Fast Twitches'\\ Width(2$^{\circ}$), Lowered(20$^{\circ}$)\\ Speed:100$^{\circ}$/s, Horizontal Wag\end{tabular} & 
\begin{tabular}[c]{@{}c@{}}`Fast Upward Twitches'\\ Width(4$^{\circ}$), Raised(35$^{\circ}$)\\ Speed:100$^{\circ}$/s, Horizontal Wag\end{tabular}  \\ \midrule
\textbf{Light} & \begin{tabular}[c]{@{}c@{}} Pink (R:255 G:135 B:200)\end{tabular} & \begin{tabular}[c]{@{}c@{}}Blue (R:0 G:50 B:240)\end{tabular} & \begin{tabular}[c]{@{}c@{}}Red (R:230 G:0 B:20)\end{tabular} & \begin{tabular}[c]{@{}c@{}}Grey (R:185 G:185 B:185)\end{tabular} & \begin{tabular}[c]{@{}c@{}}Yellow (R:255 G:255: B:70)\end{tabular} \\ \midrule
\textbf{Sound} & \begin{tabular}[c]{@{}c@{}}`Short Happy Chirps'\\ 600-1300Hz\\ Sine Sweep\\ 3 x 0.1s , 0.15s intervals\\ Repeated Once\end{tabular} & \begin{tabular}[c]{@{}c@{}}`Slow Downward Peep'\\ 500 - 300hz \\ Sine Sweep\\ 1 x 4s\\ Not Repeated\end{tabular} & \begin{tabular}[c]{@{}c@{}}`Low Fast Growls'\\ 80-160hz, 10-100\% Amp\\ Square Sweeps\\ 3 x 2s, 2s intervals\\ Repeated Once\end{tabular} & \begin{tabular}[c]{@{}c@{}}`Trembling Whine'\\ 800Hz, 1.8x Speed\\ Sine Wave\\ 2 x 1.5s , 1s interval\\ Tremolo: 0.25s interval\end{tabular} & \begin{tabular}[c]{@{}c@{}}`High Rising Peep'\\ 1000-1300Hz\\ Triangle Sweep\\ 1 x 1.5s\\ Repeated Once\end{tabular}  \\ \midrule
\textbf{Emoji} & \includegraphics[width = 3 cm, height = 1.9cm]{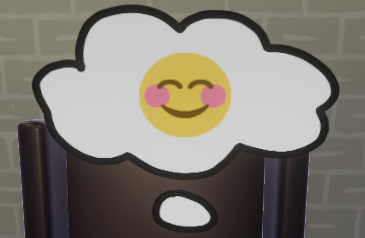} & \includegraphics[width = 3 cm, height = 1.9cm]{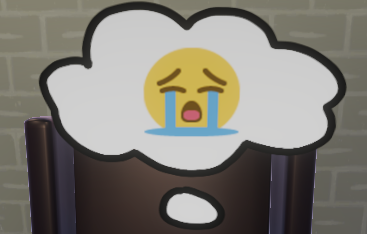} & \includegraphics[width = 3 cm, height = 1.9cm]{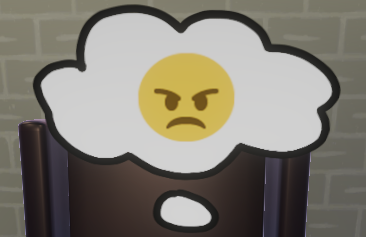} & \includegraphics[width = 3 cm, height = 1.9cm]{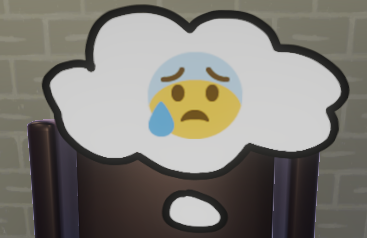} & \includegraphics[width = 3 cm, height = 1.9cm]{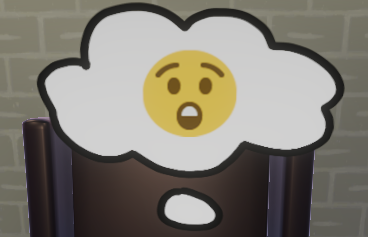} \\ \midrule
\textbf{Text} & \includegraphics[width = 3 cm, height = 0.9cm]{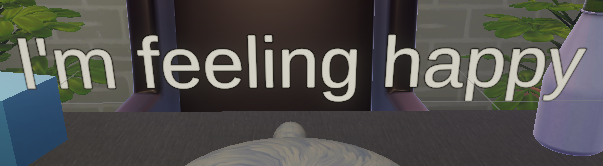} & \includegraphics[width = 3 cm, height = 0.9cm]{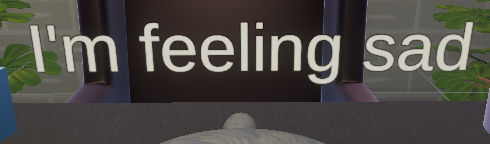} & \includegraphics[width = 3 cm, height = 0.9cm]{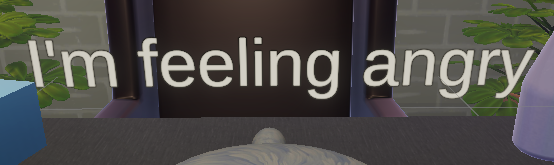} & \includegraphics[width = 3 cm, height = 0.9cm]{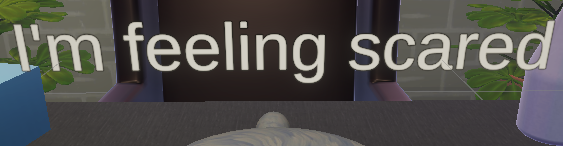} & \includegraphics[width = 2 cm, height = 0.9cm]{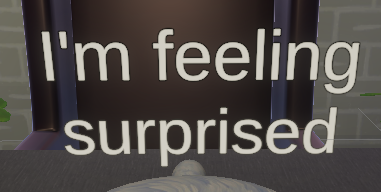} \\ \midrule
\begin{tabular}[c]{@{}c@{}}\textbf{Speech}\end{tabular} & "I'm Feeling Happy" & "I'm Feeling Sad" & "I'm Feeling Angry" & "I'm Feeling Scared" & "I'm Feeling Surprised" \\ \bottomrule
\end{tabular}
}
\label{tab:modalities}
\end{table*}

\noindent\textbf{Stage 1 - Pre-Study Procedure:}
Participants were briefed, read the information sheet and signed a consent form. 
They indicated age, gender, prior experience with VR and social robots, then completed the NARS.
The robot was then turned on briefly to familiarise participants with its current features.
They were fitted with the VR headset to complete calibration and enter the VE for training.
During training, participants practised the procedure of placing their hand on the robot to receive emotion expressions and filling in rating scales (Fig. \ref{fig:VE}).
Once ready, they pressed a button to begin.

\noindent\textbf{Stage 2 - Study Procedure:}
The stage took place in VR in seven rounds (one per modality) ordered by Latin Square. 
Five emotions were expressed in a random order per round. 
Expressions are shown in Tab. \ref{tab:modalities} and discussed in Sec. \ref{sec:modality_design}.
Participants triggered expressions by placing their hand on the robot. 
Expression were displayed for seven seconds (informed by pilot testing), during which time participants were free to touch the robot as they wished before the rating scales appeared.
Participants selected which of the five emotions they thought was expressed, or `None' if they couldn't recognise it.
They then rated how effective the expression was in terms of clarity on a five-point Likert scale from `Very Ineffective' to `Very Effective' and how empathetic, or emotionally close, they felt with the robot on a five-point scale from `Not at all Empathetic' to `Very Empathetic'.
Finally, they pressed `Confirm' and placed their hand on the robot for the next expression.
After each modality, participants could take a break and then press a button for the next round.
Participants stayed in VR throughout for immersion but were advised they could remove the headset at any time if they wished.

\noindent\textbf{Stage 3 - Post-Study Procedure:}
Participants then removed the headset and filled out the IPQ, SUS, and ranked modalities in order of preference, giving rationale as free text.
They reported prior animal experience and how this impacted their preferences, then if they had a preference for pet-like modalities (\textit{face}, \textit{sound}, \textit{tail}), or other modalities (\textit{text}, \textit{emoji}, \textit{speech}, \textit{light}), and why.

\noindent\textbf{Stage 4 - Participatory Prototyping:}
Participants re-entered the VE and were guided on how to toggle modalities on and off to design their own multimodal expressions (Fig. \ref{fig:VE}). 
While somewhat preliminary, as participants created two designs (happy and sad) using three or more modalities, this provided valuable insights into how modality preferences may change for multimodal expressions.
Participants were asked to discuss the rationale for their designs in a short interview to discuss their design rationale, then were paid, debriefed, and could ask any further questions.
Software-generated transcripts were corrected and anonymised for qualitative analysis.

\subsection{Emotion Expression Modality Design}
\label{sec:modality_design}

Seven modalities drawn from prior work and online communication were prototyped in the VE (see Tab.\ref{tab:modalities}). Source code for the VE and modalities is available to facilitate  replicability\footnote{Code: github.com/Torquoal/VR\_Zoomorphic\_Robot\_Expressions}.
The design and rationale for each modality is discussed below.  
\par
\noindent\textbf{Face:}
We drew inspiration from prior work exploring facial expressions and social robotics~\cite{Ghafurian2022, Young2007, Macdonald2019} to create five simple expressions using a DecalProjector GameObject.

\noindent\textbf{Tail:}
We designed five affective tail movements by varying the pitch, yaw, rotational speed and range of a geometric `Tail' GameObject. These movements aligned with the position of our five basic emotions on the Circumplex model~\cite{Russell1980}, drawn from prior work~\cite{Singh2013} mapping tail movements to valence and arousal, e.g., higher speed conveys greater arousal/valence, vertical wagging conveys negative valence compared to horizontal wagging. 

\noindent\textbf{Light:} A modality prominently used in prior work ~\cite{Ghafurian2022, Song2017, Thompson2024}. We consulted a cross-cultural study~\cite{Jonauskaite2019} to select colour-emotion associations. 
To address the lack of a distinct colour for happiness, we chose a similar emotion (love) with a defined colour (pink).
Conveyed with a Spotlight GameObject under the robot (Fig. \ref{fig:teaser}).

\noindent\textbf{Sound:}
As it was challenging to find discrete animal sounds for five emotions, we built upon prior work~\cite{Loffler2018, Macdonald2019} and pilot testing, generating five tones~\cite{onlinetone}, displayed via Audio Source GameObject.

\noindent\textbf{Emoji:}
We chose unicode emojis~\cite{unicode} due to their ubiquity in mediating online communication, shown above the robot in a thought bubble for clarity using a Canvas GameObject. 

\noindent\textbf{Text and Text-to-Speech (TTS):}
Alongside more interpretive modalities, we also explored messages expressed via two commonly used modalities from online communication and social media, written text (displayed via a Canvas above the robot) and TTS (generated via an online tool and displayed via Audio Source~\cite{freetts}). 
We chose generic and simple sentence structures (e.g., ``I'm Feeling Happy/Sad'') to maximise simplicity and comprehension.

\section{Results}

\subsection{Negative Attitude Towards Robots scale (NARS)}

The NARS showed that our sample held similar attitudes to social robots as in prior work~\cite{Nomura2006, Nomura08, Graaf2013}, with no strong bias~\cite{syrdal2009negative}.
It features three sub-scales, S1 (max score=30), S2 (max=25) and S3 (max=15). 
The mean score for S1 was 13.6 ($\sigma$=3.38), indicating slight acceptance towards \textit{interaction with robots}. A mean S2 score of 15.0 ($\sigma$=3.75) indicated \textit{marginal concern about robots in social spaces}.
A mean S3 score of 7.53 ($\sigma$=2.21) indicated ambivalence towards \textit{emotional interactions with robots}.

\subsection{Quantitative Analysis of Participant Ratings}

This section reports the quantitative analysis of emotion recognition accuracy and subjective ratings of expression effectiveness and feelings of empathy, blocked by participant as a random variable. 
Shapiro-Wilk tests confirmed that accuracy, effectiveness and empathy data ($p<0.001$) were not normally distributed, so we used the Aligned Ranked Transform (ART) \cite{JacokWobbrockLeahFindlaterDarrenGergle2011} to allow two-factor parametric tests to be conducted.
Statistical analysis used R and the \textit{ART}, \textit{emmeans} and \textit{eff\_size} packages. 
The main effects of modality and emotions on each dependent variable are shown in Tab. \ref{tab:MainEffects}. Summary statistics and \textit{post hoc} results are shown in Figs. 6 and 7.

\begin{table}[h!]
\centering
\caption{Main effects of modality and the emotion expressed on recognition accuracy, perceived effectiveness and empathy.}
\resizebox{1\columnwidth}{!}{%
\begin{tabular}{@{}cccccccc@{}}
\toprule
 &  & \multicolumn{2}{c}{Accuracy} & \multicolumn{2}{c}{Effectiveness} & \multicolumn{2}{c}{Empathy} \\ \midrule
Factor & $df$ & $F$ & $p$ & $F$ & $p$ & $F$ & $p$ \\ \midrule
Modality & 6 & 112.8 & \textbf{\textless{}.0001} & 42.78 & \textbf{\textless{}.0001} & 40.20 & \textbf{\textless{}.0001} \\
Emotion & 4 & 51.56 & \textbf{\textless{}.0001} & 0.616 & 0.651 & 1.564 & 0.182 \\
Modality:Emotion & 24 & 12.96 & \textbf{\textless{}.0001} & 1.284 & 0.163 & 1.312 & 0.144 \\ \bottomrule
\end{tabular}
}
\label{tab:MainEffects}
\end{table}

\begin{figure}[h!]
\centering
\includegraphics[width=.95\columnwidth]{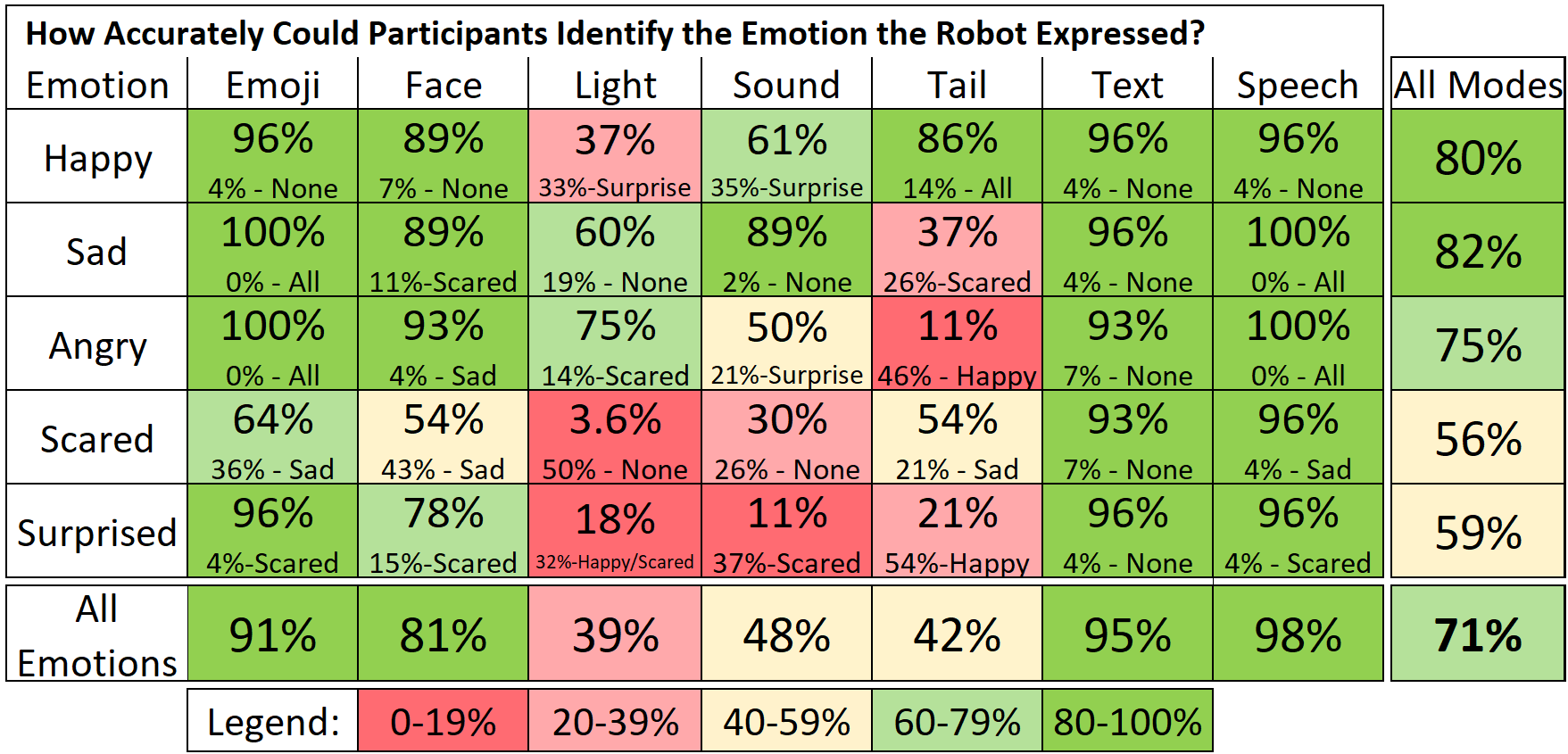}
\\
\includegraphics[width=.95\columnwidth]{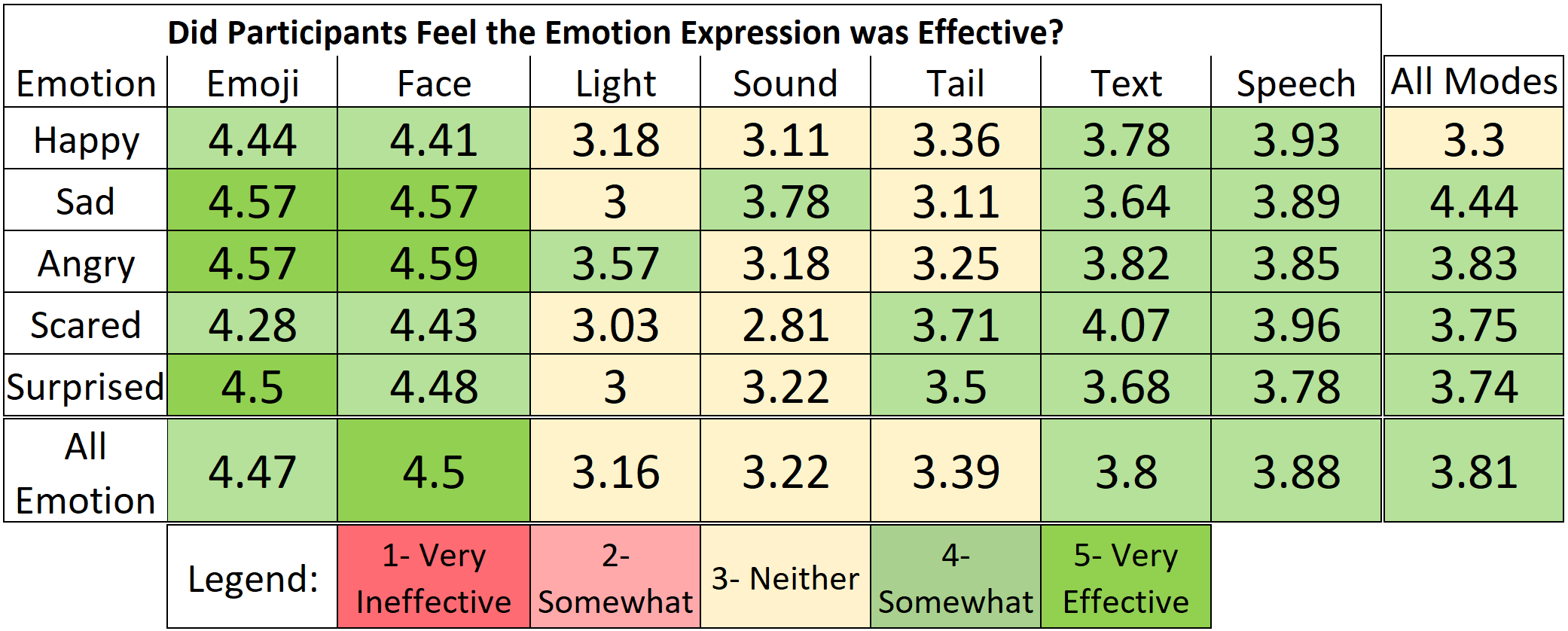}
\\
\includegraphics[width=.95\columnwidth]{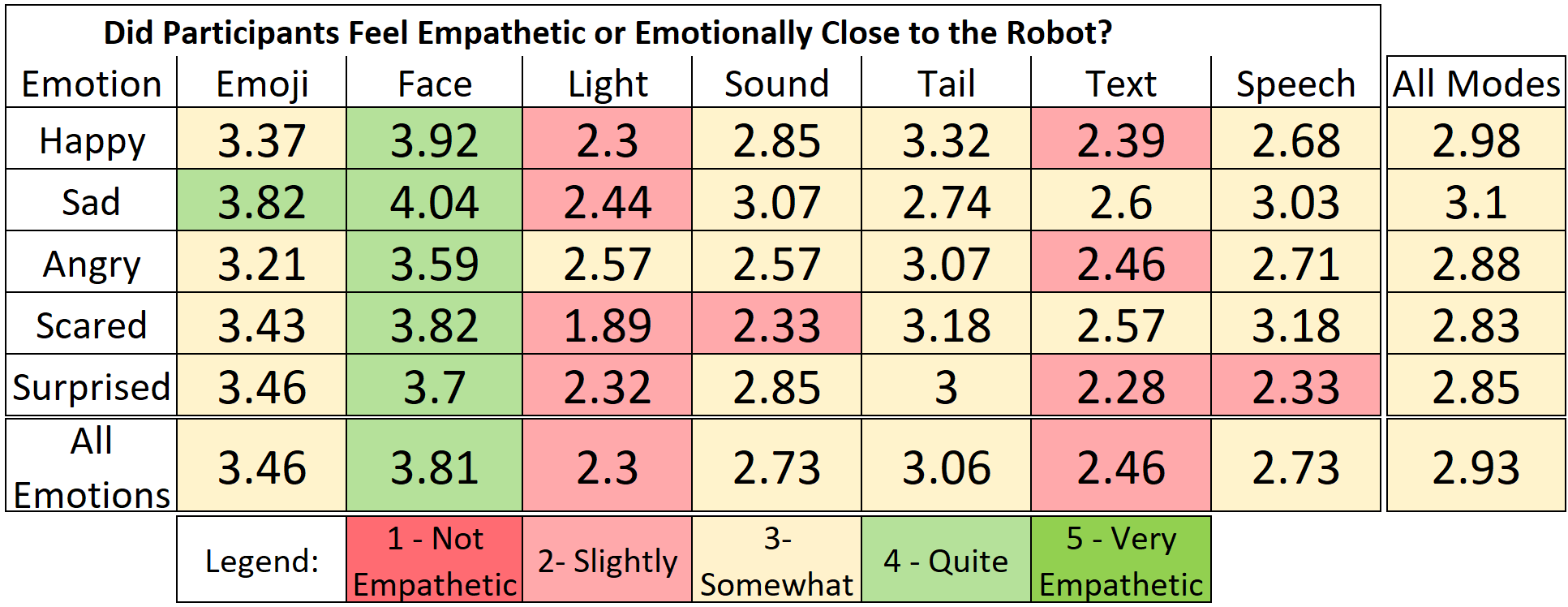}
\caption{Recognition accuracy, perceived effectiveness and empathetic response for each modality and emotion combination.}
    \label{summary_tables}
\end{figure}

\noindent\textbf{Impact on Emotion Recognition Accuracy:}
Overall emotion recognition accuracy was 71\% (Fig. \ref{summary_tables}).
Explicit modalities from online communication, \textit{emoji}, \textit{text} and \textit{speech}, were recognised \(>\)90\% of the time across emotions but not perfectly, perhaps reflecting participant input or memory error.
These were followed by \textit{facial} expressions (81\%), while 
\textit{Light}, \textit{sound} and \textit{tail} achieved lower accuracy (\(<\)48\%).
\textit{Light} was most divisive. Blue and red achieved 60\% and 75\% respectively, but grey light only 3.6\%  as half of participants associated it with no emotion.
\textit{Happiness} and \textit{sadness} were most recognisable across modalities (\(>=\)80\%) followed by \textit{anger} (75\%), \textit{surprised} (59\%) and \textit{scared} (56\%). 
\par
\begin{figure*}[h!]
\centering
\includegraphics[width=.95\textwidth]{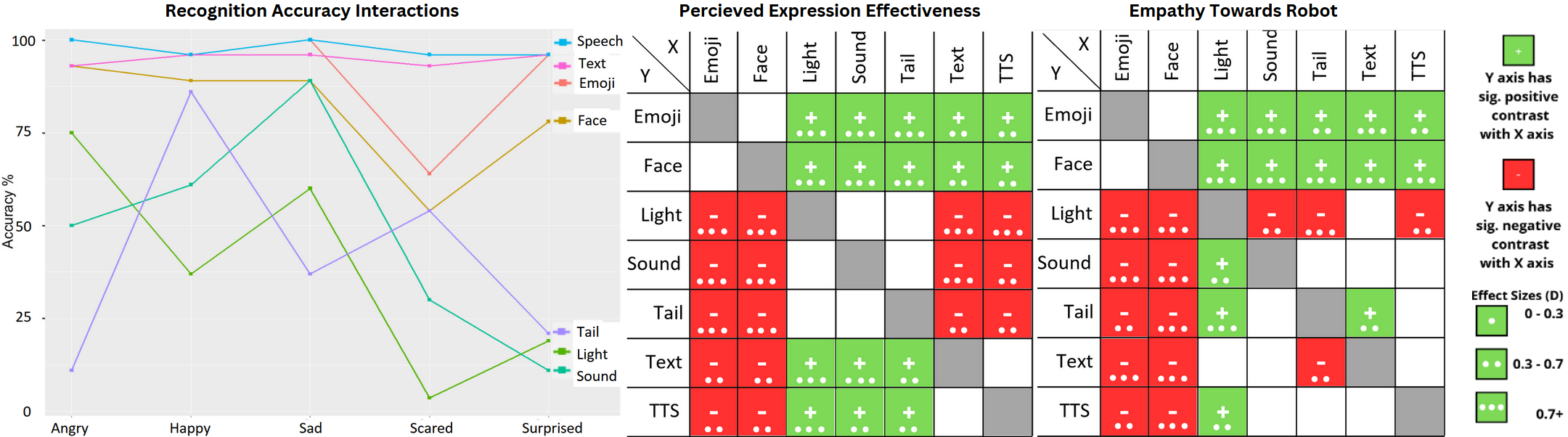}
\caption{Visualisations of post hoc accuracy interactions and pairwise contrasts for effectiveness and empathy ratings.}
    \label{post_hoc_vis}
\end{figure*}

A two-factor ANOVA found both modality and emotion had a main effect on accuracy and a significant interaction ($p<0.0001$, Tab. \ref{tab:MainEffects}).
\textit{Post hoc} pairwise t-testing with Tukey P-value adjustment ($\alpha = 0.05$) investigated contrasts between modalities, blocked by emotion and participant.
\textit{Emoji}, \textit{face}, \textit{text} and \textit{TTS} were recognised more than \textit{light}, \textit{sound} and \textit{tail} ($p<0.05$) with large effect sizes ($d>0.8$).
\textit{Text} and \textit{TTS} were recognised more than \textit{face} ($p<0.05$), with medium effect sizes ($0.3>d>0.8$).
Testing between emotions, blocked by modality and participant, found that \textit{angry}, \textit{happy} and \textit{sad} expressions were more recognisable ($p<0.0001$) than \textit{scared} or \textit{surprised}, with medium effect sizes. Several significant interactions were not accounted for by these contrasts (Fig. \ref{post_hoc_vis}).
\textit{Happy} \textit{tail} expressions were more recognisable than \textit{angry}, \textit{sad} and \textit{surprise} ($p<0.0001$), as participants often assumed tail movements were happy. 
When expressed by \textit{light}, \textit{anger} was more recognisable than \textit{happiness} ($p<0.0001$).
\textit{Sad sound} was more recognisable than \textit{anger} ($p<0.0001$).

\noindent\textbf{Impact on Perceived Effectiveness of Emotion Expression:}
Participants rated how effectively they felt that emotion was clearly displayed from `Very Ineffective' (re-coded to 1) to `Very Effective' (5).
Mean effectiveness was 3.81 with a median of `Somewhat Effective' (Fig. \ref{summary_tables}). 
The most accurate modes, \textit{text} and \textit{speech}, were only considered `Somewhat Effective', potentially due to their less emotive nature. 
\textit{Face} and \textit{emoji} were considered effective, while less recognisable modes, \textit{light}, \textit{sound} and \textit{tail}, were less effective.
Regarding emotions, \textit{happy} expressions were less effective on average, despite being recognisable.
Modality had a significant main effect on effectiveness, while emotion had no impact.
\textit{Post hoc} testing found several contrasts between modalities: \textit{emoji} and \textit{face} had large or medium positive contrasts with every other modality (Fig. \ref{post_hoc_vis}).
\textit{Text} and \textit{TTS} were perceived as more effective than \textit{light}, \textit{sound} and \textit{tail} ($p<0.0001$), with medium to large effects.

\noindent\textbf{Impact on Perceived Empathy Toward the Robot:}
Participants rated how empathetic they felt toward the robot during expressions, from `Not Empathetic' (1) to `Very Empathetic' (5).
The mean rating was 2.93, with a median of `Somewhat Empathetic' (Fig. \ref{summary_tables}).
Participants felt most empathetic during \textit{facial} expressions, then \textit{emoji}.
\textit{Sound}, \textit{tail} and \textit{speech} had middling ratings (2.7-3.1) while \textit{text} and \textit{light} were rated lowest (>2.5).
Ratings were similar across emotions.
15 participants were also observed engaging in additional touch interactions with the robot, such as stroking and scratching. 
Modality had a main effect on empathy, while emotion did not.
\textit{Post hoc} testing identified several significant contrasts ($p<0.05$).
\textit{Emoji} and \textit{face} elicited higher ratings than all others, with mostly large effect sizes (Fig. \ref{post_hoc_vis}). 
\textit{Light} elicited less empathy than \textit{sound}, \textit{TTS} and \textit{tail}. \textit{Tail} contrasted positively with \textit{text}.

\subsection{Modality Rankings} 
\begin{figure}[ht!]
\centering
    \includegraphics[width=0.48\textwidth]{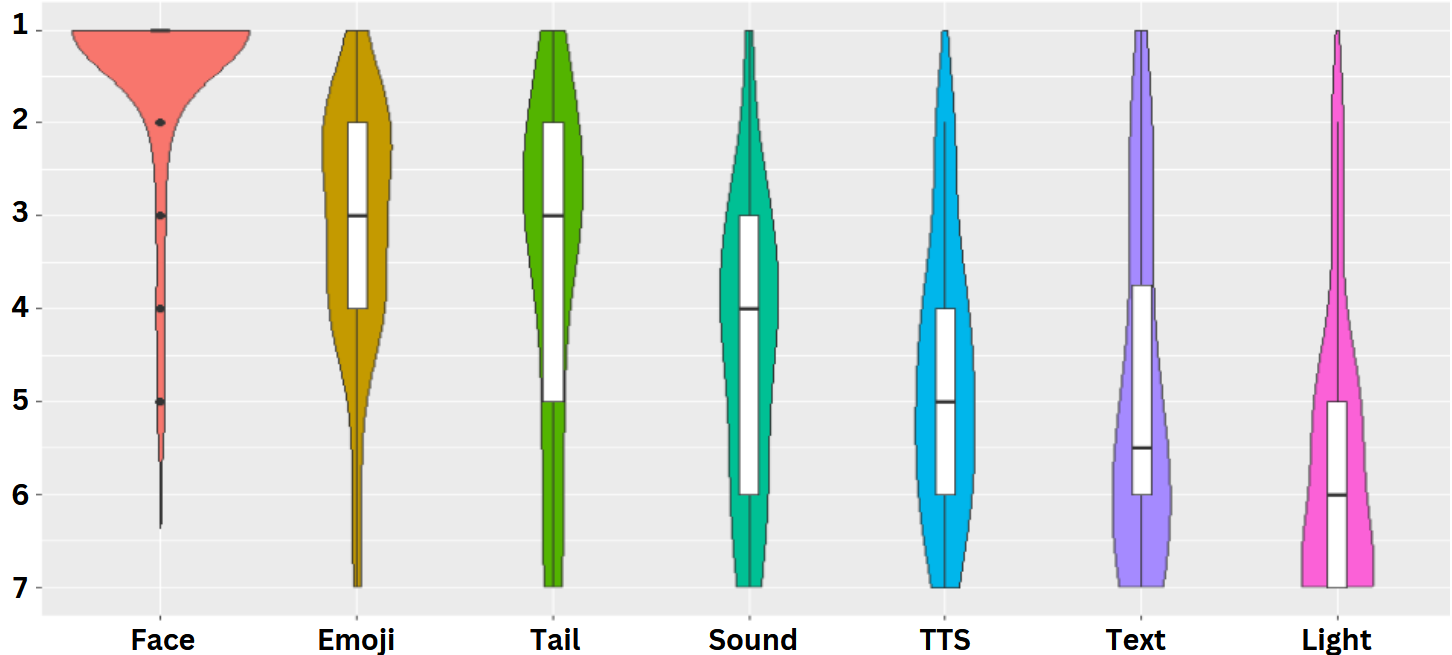}
\\
\caption{Order, distribution of participant modality rankings.}
    \label{fig:Modality_Ranks}
\end{figure}

After exiting the VE, participants ranked modalities in order of preference (see Fig. \ref{fig:Modality_Ranks}).
Rankings aligned roughly with effectiveness and empathy ratings. 
\textit{Face} had a median rank of 1st, as well as the highest average effectiveness and empathy. 
\textit{Emoji} and \textit{tail} were ranked 3rd and \textit{sound} 5th.
Tail had a median rank of 5.5 and \textit{light} was ranked 6th.
Significant differences in modality rankings were found via a Kruskal Wallis test ($\chi^2=86.0$, $df=6$, $p<0.0001$), followed by a \textit{post hoc} pairwise Dunn test with Bonferroni correction.
\textit{Face} was significantly preferred ($p<0.05$) to every other modality, while \textit{emoji} positively contrasted \textit{light} and \textit{text}. 
Finally, \textit{tail} was significantly preferred to \textit{light}.

\subsection{Presence (IPQ) and Usability (SUS) Scales}

Participants completed the IPQ to assess feelings of presence in the VE.
The mean score was 4.51 ($\sigma$=1.94), indicating `excellent' levels of presence by Melo et al.'s adjective descriptions ~\cite{Melo2023}.
The Spatial Presence sub-scale had a `very good' mean value of 5.02 ($\sigma$=0.74).
Attention and involvement were `satisfactory', with a mean value of 4.28 ($\sigma$=0.77).
Finally, Experienced Realism in the VE was `very good', with a mean value of 3.75 ($\sigma$=0.83). The SUS was then administered to check if the usability of the VE and 3DUI may have negatively impacted participants. The mean SUS value we observed was 83.3 ($\sigma$=12.3), classified as `excellent'~\cite{Bangor2009}.


\subsection{Participatory Prototyping}
\label{sec:prototyping}

\begin{table}[h]
\caption{Prevalence of modalities and modality combinations used in participatory prototyping of multimodal happy (left) and sad (right) expressions. Prevalence of modalities is bolded.}
\label{prototypemodalities}
\resizebox{.48\textwidth}{!}{%
\begin{tabular}{|lccccccc|}
\hline
\multicolumn{8}{|c|}{\textbf{Happy / Sad Prototype Modality Prevalence}} \\ 
\multicolumn{1}{|l|}{} & \multicolumn{1}{l|}{\textit{Emoji}} & \multicolumn{1}{l|}{\textit{Face}} & \multicolumn{1}{l|}{\textit{Light}} & \multicolumn{1}{l|}{\textit{Sound}} & \multicolumn{1}{l|}{\textit{Tail}} & \multicolumn{1}{l|}{\textit{Text}} & \multicolumn{1}{l|}{\textit{TTS}} \\ \hline
\multicolumn{1}{|l|}{\textit{Emoji}} & \multicolumn{1}{c|}{\textbf{13 / 12}} & \multicolumn{1}{c|}{12 / 11} & \multicolumn{1}{c|}{11 / 4} & \multicolumn{1}{c|}{2 / 9} & \multicolumn{1}{c|}{11 / 7} & \multicolumn{1}{c|}{2 / 1} & 0 / 0 \\ \hline
\multicolumn{1}{|l|}{\textit{Face}} & \multicolumn{1}{c|}{12 / 11} & \multicolumn{1}{c|}{\textbf{27 / 25}} & \multicolumn{1}{c|}{15 / 11} & \multicolumn{1}{c|}{10 / 19} & \multicolumn{1}{c|}{26 / 21} & \multicolumn{1}{c|}{7 / 4} & 1 / 0 \\ \hline
\multicolumn{1}{|l|}{\textit{Light}} & \multicolumn{1}{c|}{11 / 4} & \multicolumn{1}{c|}{15 / 11} & \multicolumn{1}{c|}{\textbf{17 / 15}} & \multicolumn{1}{c|}{4 / 8} & \multicolumn{1}{c|}{15 / 11} & \multicolumn{1}{c|}{5 / 2} & 0 / 0 \\ \hline
\multicolumn{1}{|l|}{\textit{Sound}} & \multicolumn{1}{c|}{2 / 9} & \multicolumn{1}{c|}{10 / 19} & \multicolumn{1}{c|}{4 / 8} & \multicolumn{1}{c|}{\textbf{10 / 23}} & \multicolumn{1}{c|}{10 / 17} & \multicolumn{1}{c|}{2 / 2} & 0 / 0 \\ \hline
\multicolumn{1}{|l|}{\textit{Tail}} & \multicolumn{1}{c|}{11 / 7} & \multicolumn{1}{c|}{26 / 21} & \multicolumn{1}{c|}{15 / 11} & \multicolumn{1}{c|}{10 / 17} & \multicolumn{1}{c|}{\textbf{26 / 23}} & \multicolumn{1}{c|}{6 / 5} & 0 / 0 \\ \hline
\multicolumn{1}{|l|}{\textit{Text}} & \multicolumn{1}{c|}{2 / 1} & \multicolumn{1}{c|}{7 / 4} & \multicolumn{1}{c|}{5 / 2} & \multicolumn{1}{c|}{2 / 2} & \multicolumn{1}{c|}{6 / 5} & \multicolumn{1}{c|}{\textbf{7 / 5}} & 0 / 0 \\ \hline
\multicolumn{1}{|l|}{\textit{TTS}} & \multicolumn{1}{c|}{0 / 0} & \multicolumn{1}{c|}{1 / 0} & \multicolumn{1}{c|}{0 / 0} & \multicolumn{1}{c|}{0 / 0} & \multicolumn{1}{c|}{0 / 0} & \multicolumn{1}{c|}{0 / 0} & \textbf{1 / 0} \\ \hline
\end{tabular}
}
\end{table}

\begin{figure}[ht!]
    \centering
         \includegraphics[width=0.17\textwidth]{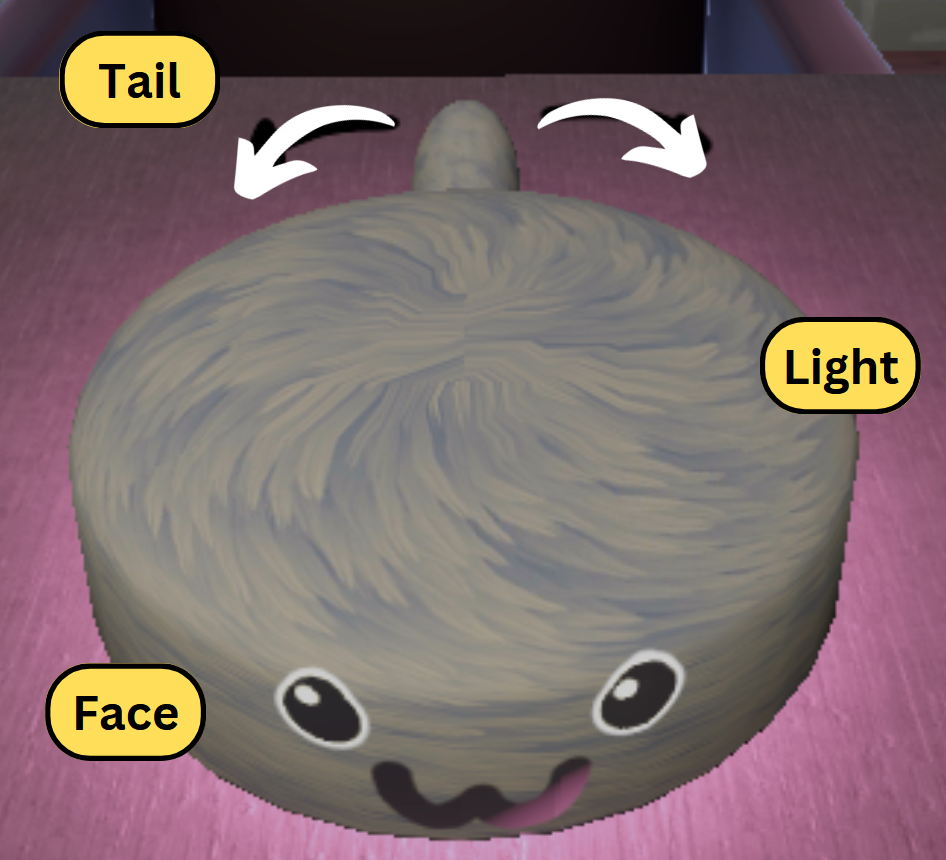}
         \hspace{1cm}
         \includegraphics[width=0.17\textwidth]{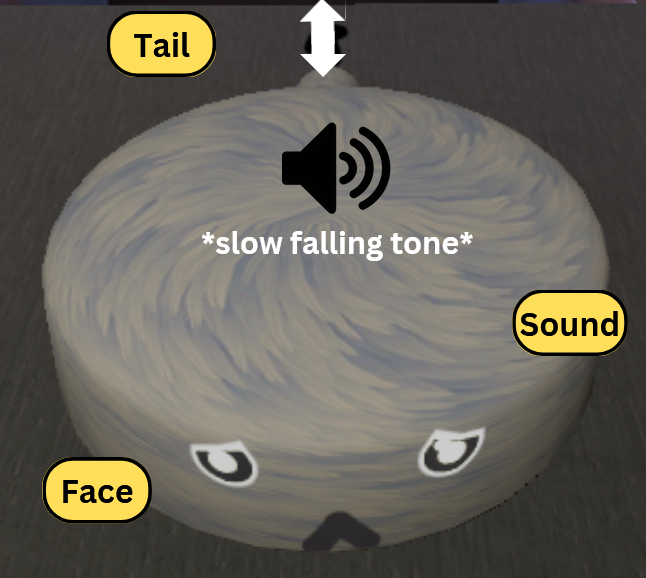}
    \caption{The most prevalent happy and sad multimodal prototypes.}

    \label{fig:ModelledPrototypes}
\end{figure}

Participants prototyped multimodal expressions for two core emotions with opposite valence, happiness and sadness, using at least three modalities inside the VE.
Tab. \ref{prototypemodalities} shows the total prevalence of modalities and how often they were used in combination.

\noindent\textbf{Happy Multimodal Prototypes}
To express happiness, 15 participants used three modalities, nine used four, and four used five. 
The most prevalent combination was \textit{face}, \textit{tail}, \textit{light}, used 15 times (Fig. \ref{fig:ModelledPrototypes}).
The most prevalent modalities, \textit{face} (27 times) and \textit{tail} (26), were always combined. 
Although second in modality rankings, \textit{emoji} were only used 13 times, while \textit{light} was ranked last but used 17 times.
The rationale for this is explored in Sec. \ref{sec:QualitativeAnalysis}.
\textit{Light} and \textit{emoji} were used together 11 times and closely coupled with \textit{face} and \textit{tail}.
\textit{Sound} and \textit{text} were used 10 and 7 times respectively, always alongside \textit{face} and \textit{tail}.
Only one person used \textit{speech}.

\noindent\textbf{Sad Multimodal Prototypes}
To express sadness, 13 participants used three modalities, 14 used four and one used five. 
The most prevalent combination (16), was \textit{face}, \textit{tail} and \textit{sound}.
Again \textit{face} (25) and \textit{tail} (23) were most prominent and tightly coupled (21).
\textit{Sound} was more prevalent in sad designs (21).
\textit{Light} (15), \textit{emoji} (12) and \textit{text} (5) were all used less often than for happy prototypes, as compliments to the \textit{face} and \textit{tail}.
\textit{Speech} was unused. 


\subsection{Thematic Analysis of Qualitative Feedback}
\label{sec:QualitativeAnalysis}

\subsubsection{Qualitative Analysis Methodology}

We applied thematic analysis to qualitative survey and interview data~\cite{braun2013successful, Braun2006}.
Two researchers familiarised themselves with the data, then conducted an initial inductive coding pass.
After coding an identical 50\% data subset, the researchers met to normalise codes.
With this initial codebook, both researchers separately re-coded the entire data set, making alterations as required, then met to synthesise codes\footnote{We took a qualitative approach to validity on advice from the creators of thematic analysis to not calculate quantitative Inter-Coder Reliability~\cite[p.278]{braun2013successful}, acknowledging the influence of the researcher on the process.}.
Axial coding was then used to form categories and identify relationships, informing four descriptive themes. 
For the final codebook of 32 codes, see the supplemental materials.

\subsubsection{Theme 1: Interpretability and Evocativeness}

Participants strongly valued modalities that were easy to interpret and emotionally expressive, while evoking real animals.

\noindent\textbf{Interpretable:}
Many participants' rankings ``\textit{most closely follows the need for clarity}'' (Participant 1/P1), highlighting the \textit{face}, \textit{emoji}, \textit{text} and \textit{TTS}. 
Less clear modalities (\textit{light}, \textit{sound}, \textit{tail}) resulted in ``\textit{guessing games}'' (P23), with P25 ``\textit{confused about intended emotion in the case of sounds and coloured light}''.
Interpretability was often grounded in familiarity: ``\textit{facial expressions and emojis were easy to identify as I see them commonly}'' (P13).
Some modalities may require learning: ``\textit{I wasn't sure about the noises and what they meant but if I had prior learning it would have been helpful}'' (P2), but could be effective: ``\textit{[sound, light and tail] ... once you learned them they would create a better emotional connection}'' (P15).

\noindent\textbf{Evocative:}
Participants preferred modalities that evoked pets to feel ``\textit{more connected to [Qoobo]}'' (P11), such as P22: ``\textit{I ranked [modalities] higher that I associated with a dog}''.
\textit{Sound}, \textit{tail} and particularly \textit{face} were mentioned: ``\textit{I preferred the facial expression as it allowed me to consider the robot as if it was real and was actually expressing the emotion}'' (P28).
\textit{Text} and \textit{TTS} were less emotive: ``\textit{facial expressions and tail movement made the robot more life-like and relatable than text}'' (P19), which harmed clarity: ``\textit{confusion created when the speech doesn’t match the emotion}'' (P1).
P12 summarised this tension:``\textit{balance between "emotive enough" and "not too artificial seeming". So tail movement was moderately emotive and seemed quite realistic coming from Qoobo, so I liked it the best. Facial expressions were really emotive but slightly artificial.}''

\subsubsection{Theme 2: Pet-Like Versus Other Modalities}

Participants expanded on tension between pet-like (\textit{face}, \textit{sound}, \textit{tail}) and non-pet modalities (\textit{emoji}, \textit{text}, \textit{light}, \textit{TTS}). 
18 preferred pet-like, 1 preferred non-pet and 9 preferred combinations.

\noindent\textbf{Pet-Like Modalities}:
The familiarity of pet-like modes motivated preference.
P27 felt without pet-like expressions they may ``\textit{feel scared when I play with these robot-based animals}''.
They allowed instinctual understanding: ``\textit{I don't need to think what the robot is trying to say}'' (P22), but some worried about clarity: ``\textit{tail movements and sounds are good in the real world but if we put a comparison then emojis, speech and text help us understand}'' (P23).

\noindent\textbf{Non-Pet Modalities:}
Non-pet modalities could be unemotive or unimmersive: ``\textit{created a bit of distance and almost reminded me of being in a virtual space}'' (P12). 
Their clarity was highlighted, however, as P11 felt that ``\textit{speech and text can be added to the robots as it would be easier for the human to understand their emotions}'', while P7 wrote: ``\textit{adding [...] modes such as emoji will help to understand the emotion more. Just like [...] with phone messages}''.

\subsubsection{Theme 3: Expectations Driven by Animal Experience}
Participant impressions were contextualised by prior animal experience.
Some participants with less experience felt they lacked context: ``\textit{I might have misinterpreted the Qoobo's feelings via anthropomorphism, I might not be as familiar with reacting to animal gestures}'' (P2).
Animal experience formed the basis for interpretation for other participants.
It could raise expectations: ``\textit{the robot need to be a bit more complex}'' (P18), or make human-like modes dissonant: ``\textit{strange and unrealistic coming from an animal-ish creature}'' (P12).
Some modalities evoked prior experiences, with P8 and P15 highlighting that \textit{tail} movement was ``\textit{like}'' or ``\textit{the same}'' as an animal.
This could heighten emotional connection: ``\textit{feel empathetic towards the robot as it looks and behaves a bit like a pet}'' (P13).

\subsubsection{Theme 4: Multimodal Designs - Naturalistic to Holistic}

Most grounded their multimodal designs in pet-like modalities for to ``\textit{made me feel connected}'' (P11); ``\textit{like a pet or person}'' (P4).
Some found these interpretable: (``\textit{naturalistic and familiar}'' P26), but others sought clarity with non-pet modes.
P14 added emoji: ``\textit{you could easily check what the emotion is}'', while P1 was a ``\textit{fan of the first person text because I get that really quick}''.
Others combined visuals and audio for emphasis: ``\textit{auditory feedback that tells you how they're feeling like as an accompaniment to the visual gestures. It's just amplification}'' (P2).
Despite low ratings, \textit{light} was used to compliment many designs: ``\textit{colours are really good at setting the mood}'' (P11).
Overall, participants used evocative expressions, then augmented those designs with a mixture of pet-like and non-pet elements to strengthen and clarify these expressions.

\noindent\textbf{Emotion-Specific Modalities:}
As an aside, the interpretation of \textit{light} and \textit{sound} varied between happy or sad expressions. For P5 ``\textit{[pink] light for the happy one made sense, for the sad [...] it seemed a little bit alarming, borderline creepy}'', while P22 said ``\textit{blue [...] could even be more soothing}''.
\textit{Sound} was used often in sad prototypes, with P27 noting: ``\textit{[its] like somebody's crying so the sound really let you know what they're always feeling}'', and P28 concurred: ``\textit{you get this emotional connection to the robot}''.

\section{Discussion}

\subsection{Limitations and Future Work}
\label{limitations}

While precedent exists for using VR to assess affective displays and prototype complex systems (Sec. \ref{prioruse}), and we took steps to reduce the gap between VE and real-world experiences (Sec. \ref{VE}), the applicability of our approach is limited without future direct comparison.
We position this work to support the obstructive future undertaking of hardware or AR prototyping with real zoomorphic robots by recommending promising modalities and implementations (Sec. \ref{sec:disc_s3}). Future work should include more in-depth participatory prototyping of zoomorphic robot affective design, given the valuable insights gained from our early efforts. 
\par
We limited our approach to subjective measures in this work to focus on user perspectives and perceptions of robot emotion expression. Future work exploring physiological measures would, however, be valuable to objectively measure users' affective processing or state during interaction with zoomorphic robots~\cite{barathi_affect_2020, Wilson2015}. 
\par
In this study participants performed undirected touch to trigger emotion expressions, to evoke real interaction with zoomorphic robots. We did not have touch vary between emotions, to avoid priming, but future work should explore if there is an overall priming effect of using touch versus other actuation methods.
The results of this evaluation are inherently limited to the designs we used for modality/emotion combinations. The rationale for our designs is discussed in Sec. \ref{sec:modality_design}.
Finally, our findings should be understood in the context of the sample's relatively narrow age range (21-40).
Future work is also needed to generalise our findings to children or users with lower technology affinity, as over half of our sample had used VR `a few times' or more. 
Additionally, ratification of our recommendations for the established user demographic of older users ~\cite{Hudson2020, Shibata2009} would inform both current care and future domestic use.

\subsection{Comparing Emotion Expression Modalities}

\noindent\textbf{The Importance of a Face:}
\label{sec:disc_s1}
The \textit{face} stood out across the evaluation: highest-ranked, most used in prototypes, rated most effective, most empathetic and recognised 81\% of the time. Similarly, \textit{Emoji} also performed well.
Prior work shows that human facial expressions are key in determining emotion ~\cite{de2015perception} and that they evoke shared emotion in observers~\cite{keltner2000facial}.
Although we used simple cartoon-like non-human faces, we found they conveyed similar benefits results, supporting similar findings in non-robot studies~\cite{zhao2019event}.  
While prior work has used zoomorphic robots with facial expressions~\cite{Ghafurian2022, Shibata2009}, we demonstrate through direct comparative evaluation that this modality should be prioritised. 
\par
\noindent\textbf{Performance of Established Modes:}
We evaluated several modalities from the field, including \textit{light}, \textit{sound} and motion (\textit{tail}).
For \textit{light}, blue (sad) and red (angry) were recognisable, as in prior work ~\cite{Song2017, Loffler2018}, but happiness continues to be a challenge. 
Others have found green ineffective~\cite{Song2017}, or successfully used yellow or orange~\cite{Loffler2018, Ghafurian2022}. 
Our choice, pink, only achieved 37\% accuracy, while we used yellow for surprise. Only 33\% misidentified yellow as happiness, making it inappropriate for either emotion. 
Given our results and the cultural differences in colour-emotion associations~\cite{Jonauskaite2019}, we recommend designers explore personalisation to allow robot users to pre-select colour associations.
\par
For \textit{sound}, a falling beep for sadness was effective and emotive, echoing prior work~\cite{Loffler2018, Song2017}.
Rising beeps for happiness~\cite{Loffler2018} were somewhat recognisable, but other tones were less interpretable or too `artificial'. 
Future designers could explore alternate approaches using natural or animal-like sounds \cite{Thompson2024}, such as modulating them with the patterns of sadness and happiness we found effective.
\par
We implemented five distinct \textit{tail} movements drawn from prior work~\cite{Singh2013} but found participants often based their judgment only on tail height, confusing raised tail movements (anger/surprise) with happiness, and downward movements (scared/sad) with each other.
This may indicate these movements did not align with user expectations of real animal behaviour, perhaps due to limited or contradictory experiences. Generally, it motivates future work exploring how tail movements should vary to express distinct emotions.
Real emotional displays are often inherently expressed through many modalities governing differing aspects of expression~\cite{keltner2017understanding}. 
We saw this in our participant's prototypes, as \textit{light} \textit{sound} and \textit{tail} were limited in isolation but were used in multimodal designs to make expressions feel more amplified, social or natural, respectively.
\par
\noindent\textbf{The Lifelessness of Text and TTS:}
Despite high accuracy and effectiveness, \textit{Text} and \textit{TTS} were disliked and rarely used in prototypes in favour of evocative modalities. 
While the tone of the \textit{TTS} was somewhat modulated based on sentiment, participants felt that both modalities consistently communicated flat affect or sadness.
Given their high clarity, future work to make these two modes more emotive would be valuable, perhaps exploring more emotive voice performances, different fonts or expressive text.
\par
\noindent\textbf{User Priorities for Affective Expression:}
Our participants exhibited several \textit{priorities} for affective expressions. 
They wished for modalities to evoke animal interactions and provoke an associated emotional response, exemplified by \textit{face}, \textit{tail} and \textit{sound}.
Expressions should be \textit{interpretable} to avoid guessing or vagueness; \textit{face}, \textit{emoji} or \textit{text} achieved this.
Finally, robots should leverage \textit{multimodal} expressions that engage users in not only overt but also \textit{ambient} ways, such as \textit{light} and \textit{sound}, altering the ambience of shared space to indicate state and provoke interaction.

\begin{figure}[h!]
    \centering
    \includegraphics[width=.45\textwidth]{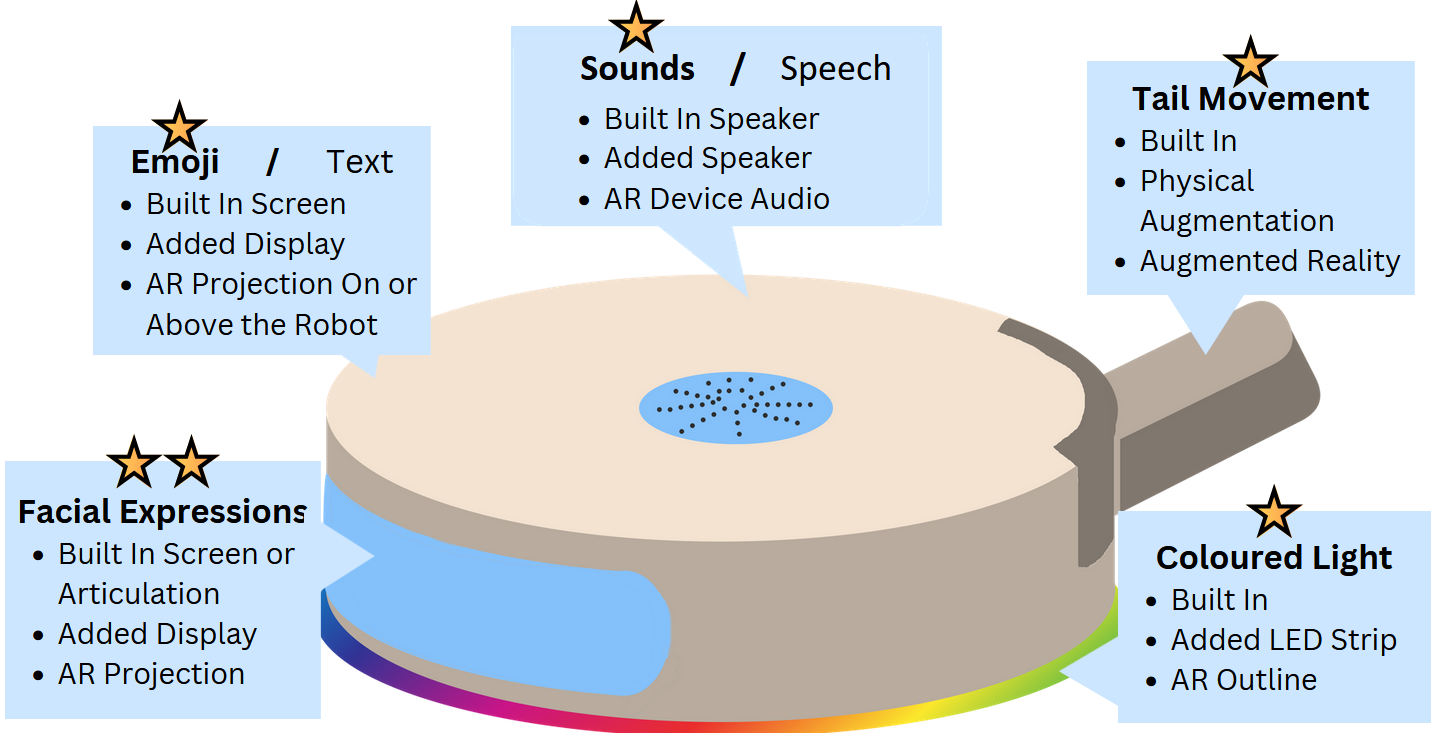}
    \caption{Implementations for emotion expression modalities, modelled on an abstract robot design. Recommended options are marked with a \textbf{star}, most recommended with \textbf{two stars}.}
    \label{fig:Implementation}
\end{figure}

\subsection{Implications of VR prototyping for HRI}
\label{sec:disc_s2}
This work highlights the potential benefits of VR prototyping for zoomorphic robot design, facilitating broad rapid exploration of affective design space and participatory prototyping.
Participants felt present in VR with the robot and engaged in physical affective touch, reducing the interaction gap between previous screen-displayed prototypes and real robots~\cite{Saldien2010, Ghafurian2022}.
VR offers practical benefits compared to physical or even AR prototyping, allowing total control of the robot's current or prospective capabilities without obstructive physical modification.
It could also facilitate rapid customisation of a robot's appearance.
Existing software allows for VR co-presence with prototyped robot interactions~\cite{abb}, but it serves industrial use cases.
We propose that future work builds on our approach to further explore affective HRI with VR prototyping. 
For example, it would be used to explore further iterations of promising modalities such as the \textit{face}, \textit{emoji} or \textit{tail}, or develop complex or evolving two-way emotional interactions between robot and user.
Overall, VR prototyping has the potential to make exploring this field faster and more accessible to varying expertise.
\par
As a by-product, these modalities may have applicability to social VR, such as VR pets, or user-controlled avatars.
Our investigation was, however, grounded in affective robot research rather than social VR (Sec. \ref{sec:modality_design}), so future work would be needed to explore this use case.
Similarly, validating the applicability of our findings to real-world use requires direct comparison.
To support this next step, we outline recommended emotion expression modalities and implementations with physical hardware or AR modification.

\subsection{Modality Recommendations and Implementations}
\label{sec:disc_s3}

\noindent\textbf{Modalities for Different Use Cases:}
Deploying expressions across different use cases may incur varied requirements.
For example, zoomorphic robots are currently used to facilitate comfort and social interaction in care settings. Here, expressions that are clear and evocative may take priority, to ensure interactions are non-distressing and understandable for users with varied capabilities. The \textit{face} and \textit{tail} should be an evocative pairing, but \textit{text} or \textit{emoji} could be added for clarity. 
By contrast, in future home use it may be more important that zoomorphic robots can engage and interact with one or more users in a shared space as a pet might, beyond controlled interactions. Our participants enabled this by using \textit{light} and \textit{sound} to allow the robot to signal its emotional state in an ambient manner. 
For all cases, if zoomorphic robots can communicate their emotional state in an effective and evocative manner it will provide a key building block for richer affective interactions.

\noindent\textbf{Recommendations for Physical Zoomorphic Robot Design:}
Four useful modalities stand out as deliverable by hardware modification (Fig. \ref{fig:Implementation}). 
\textit{Light} and \textit{sound} are more simple and transferable.
\textit{Light} can be included easily via addressable LEDs, while\textit{sound} (or speech) requires an external or inlaid speaker. 
While not off-the-shelf, prior work has demonstrated mounting emotive hardware\textit{tails} onto existing robots~\cite{Singh2013}.
\textit{Facial} expressions, while best performing, are also the most complicated to display.
Mechanical articulation is specific to each robot's design (see Miro, Fig. \ref{ZoomorphicRobots}), but a transferable option for existing robots is to mount a screen or LED panel.
Facial expressions may also clash with robots with a pre-existing unexpressive face, such as Paro, in which case \textit{emoji} could be displayed instead.
Mounting a display could still be challenging, however, due to the restrictions of each specific robot's design.

\noindent\textbf{Extending Robot Emotional with Augmented Reality:}
A possible future approach for deploying affective expressions to zoomorphic robots is via augmented reality (AR).
As seen in prior work \cite{Groechel2019}, using AR could circumvent hardware restrictions, extend robot capabilities without redesign or modification, as well as facilitate greater personalisation (Sec. \ref{sec:prototyping}).
Our VR implementation could act as a starting point for these AR interfaces, as both are experienced with similar perspectives and interactions and can be developed in similar platforms like Unity.
The applicability of some modalities could still be limited for robots with specific physical features (e.g., tail or face)), but modalities like emoji, light and sound can still be employed.
Prior work has demonstrated AR object detection for mobile robots, ensuring robots are identified in the environment~\cite{LAMBRECHT2021102178}.
However, while most smartphones can be used to enable AR, glasses or visors that would allow for seamless and unimpeded interaction with a zoomorphic robot are still specialist or hobbyist items.
While more future-facing, this approach contributes another use case for the generalised augmented reality that in turn adds value to future zoomorphic robots.

\section{Conclusion}
This work leveraged VR prototyping to conduct a broad exploration of zoomorphic robotic emotion expression.
In a mixed-methods user study, we assessed how effectively modalities conveyed affective state and evoked empathy, while qualitative feedback and participatory prototyping allowed a deeper exploration of user preferences.
The best-performing modality, facial expressions, were clear, emotive, and evocative.
Sound, coloured light and tail motion were harder to interpret but effective in multimodal designs.
Text and text-to-speech were recognisable but disliked by participants, being perceived as unnatural and unemotive.
We use these findings to recommend transferable emotion expression modalities for future zoomorphic robot prototyping, as well as physical and AR implementations. 
Furthermore, we discuss how our approach could facilitate fast and accessible exploration of this design space.

\acknowledgments{%
This work was supported by a Research Themes Award issued by SICSA, the Scottish Informatics and Computer Science Alliance.
}
\bibliographystyle{abbrv-doi}

\bibliography{references}
\end{document}